\documentclass[11pt]{article}
\usepackage{amsmath}
\usepackage{amssymb}
\usepackage{amsfonts}
\usepackage{esvect}
\usepackage{cite}
\usepackage{physics}
\usepackage{mathrsfs}
\usepackage{authblk}
\usepackage{geometry}
\usepackage{abstract}
\newtheorem{mypro}{Proposition}[section]
\title{\textbf{Extended Laplace-Runge-Lentz vectors, new family of superintegrable systems and quadratic algebras}}
\author{\large\bf Zhe Chen\footnote{zhe.chen3@uqconnect.edu.au}, 
Ian Marquette\footnote{i.marquette@uq.edu.au}, and Yao-Zhong Zhang\footnote{yzz@maths.uq.edu.au}}
\affil{\em School of Mathematics and Physics, The University of Queensland,\\ Brisbane, QLD 4072, Australia}

\begin{document}

\maketitle

\begin{abstract}
We present a useful proposition for discovering extended Laplace-Runge-Lentz vectors of certain quantum mechanical systems. We propose a new family of superintegrable systems and construct their integrals of motion. We solve these systems via separation of variables in spherical coordinates and obtain their exact energy eigenvalues and the corresponding eigenfunctions. We give the quadratic algebra relations satisfied by the integrals of motion. Remarkably these algebra relations involve the Casimir operators of certain higher rank Lie algebras in the structure constants.
\end{abstract}

\vskip.6in
\section{Introduction}
Superintegrable systems are extremely important in classical and quantum mechanics due to their rich applications. A $D$-dimensional quantum system described by the Hamiltonian 
\begin{equation}
    \hat{H}=-\sum_{i=1}^{D}\frac{\partial^2}{\partial x_i^2}+V(x_1,\cdots,x_D)
\end{equation}
is said to be integrable if there exist $D-1$ algebraically independent linear operators $I_a$ satisfying
\begin{equation}
    \comm{I_a}{\hat{H}}=0,~~~\comm{I_a}{I_b}=0,~~~a,b=1,\cdots,D-1.
\end{equation}
It is said to be superintegrable if there exist additional $k$ $(1\leq k \leq D-1)$ operators $\{J_1,\cdots,J_k \}$ commuting with $\hat{H}$
\begin{equation}
    \comm{J_i}{\hat{H}}=0,~~~i=1,\cdots,k,
\end{equation}
such that $\{\hat{H},I_1,\cdots I_{D-1},J_1,\cdots,J_{k} \}$ is algebraically independent. $J_i$'s need not commute with $I_a$'s or with each other.
\\
\\
A well-known example is the $D$-dimensional hydrogen atom (a Kepler-Coulomb system). The hydrogen atom is superintegrable  because its Hamiltonianthe commutes with the generators of Lie algebra $so(D+1)$ \cite{1,2,3}. Another famous example of superintegrable systems is the $D$-dimensional harmonic oscillator, whose Hamiltonian commutes with the generators of $su(D)$ \cite{4,5}. Superintegrable systems continue to be a very active research field. Classification of two-dimensional superintegrable systems and associated quadratic algebras were given in \cite{6}. Classification of three-dimensional superintegrable systems has not been completed, but some quadratic algebras have been obtained from their integrals of motion \cite{7,8,9,10,11}. To our knowledge, no classification has so far been available for $D$-dimensional superintegrable systems with $D>3$. 
\\
\\
The authors in \cite{RW} and \cite{KWP} independently introduced a $D$-dimensional generalization of the three-dimensional superintegrable Kepler-Coulomb system with non-central terms considered in \cite{Evan,GPS,KWP1}. This  $D$-dimensional system is maximally superintegrable with quadratic
integrals of motion. It was solved in \cite{RW} by means of separation of variables in the parabolic and spherical coordinates, and its wave functions were given in terms of orthogonal polynomials. In \cite{LMZ} the underlying symmetry algebra structure of the system was obtained and used to derive the energy spectrum algebraically.
\\
\\
In this work, we introduce a new family of superintegrable systems in $D$-dimensional space which contains the model in \cite{RW} as a special case. Our systems possess quadratic integrals of motion  and can be solved by means of separation of variables in the spherical coordinates. We give some quadratic algebra relations satisfied by these integrals. Some of the integrals are the extended Laplace-Runge-Lenz vectors which can be obtained directly from application of the proposition presented in section $2$ below.
\\
\\
This work is organized as follows. In section $2$, we give an useful proposition which demonstrates the existence of extended Laplace-Runge-Lenz vectors in certain quantum systems. In section $3$, we present new family of quantum systems in $D$-dimesnional space, and show that they are superintegrable systems by identifying their integrals of motion.
 In section 4, we solve the systems by means of separation of variables in spherical coordinates and obtain their energy eigenfunctions and eigenvalues. In section 5, we derive the underlying quadratic algebras satisfied by the integrals of motion.   Sections 5 is devoted to conclusions and comments for future work.



\section{Extended Laplace-Runge-Lenz vectors}
Let us consider the $D$-dimensional hydrogen atom
\begin{equation}\label{atom}
    \hat{H}_h=-\sum_{i=1}^{D}\frac{\partial^2}{\partial x_i^2}-\frac{\eta}{r},~~~~r=\sqrt{\sum_{i=1}^{D}x_i^2},
\end{equation}
where $\eta$ is a real parameter. $\hat{H}_h$ commutes with $D(D+1)/2$ linear operators \cite{RW}, namely
\begin{equation}
    L_{ij}=x_ip_j-x_jp_i,~~~p_i=\frac{\partial}{\partial x_i},~~~i,j=1,2,\cdots,D,
\end{equation}
and
\begin{equation}\label{glrl}
    M_i=\sum_{a=1}^{D}(p_aL_{ia}+L_{ia}p_a)+x_i\frac{\eta}{r},~~~i=1,2,\cdots,D.
\end{equation}
$L_{ij}$'s are the angular momenta, and $M_i$'s are the so-called Laplace-Runge-Lenz vectors. They satisfy following commutation relations
\begin{equation}
    \begin{aligned}
    &\comm{\hat{H}_h}{L_{ij}}=\comm{\hat{H}_h}{M_i}=0,\\
    &\comm{L_{ij}}{L_{kl}}=\delta_{jk}L_{il}+\delta_{il}L_{jk}-\delta_{ik}L_{jl}-\delta_{kl}L_{ik},\\
    &\comm{L_{ij}}{M_k}=\delta_{jk}M_i-\delta_{ik}M_j,\\
    &\comm{M_i}{M_j}=-4\hat{H}_hL_{ij}.\\
    \end{aligned}
\end{equation}

Now we consider the extension of $\hat{H}_h$, 
\begin{equation}
    \hat{H}=\hat{H}_h+V,\label{extended-H}
\end{equation}
where $V$ is some potential function to be determined later. We impose two conditions for $V$
\begin{enumerate}
    \item[(i)] $V$ is  independent of the last $m$ coordinates, i.e. \begin{equation}\label{cd1}
        [p_i,V]=\frac{\partial V}{\partial x_i}=0,~~~i=D-m+1,\cdots,D.
    \end{equation}
    \item[(ii)] $V$ is an eigenfunction of the Euler operator $\sum_{i=1}^{D}x_i\frac{\partial}{\partial x_i}=r\frac{\partial}{\partial r}$ with eigenvalue $-2$ \begin{equation}\label{cd2}
        \sum_{i=1}^{D}x_i\frac{\partial V}{\partial x_i}=-2V=r\frac{\partial V}{\partial r}.
    \end{equation}
\end{enumerate}
{}From the conditions \eqref{cd1} and \eqref{cd2}, it is not difficult to conclude $V$ is of the form
\begin{equation}\label{V}
    V=\frac{1}{\sum_{i=1}^{D-m}x_i^2}f(\phi_1,\phi_2,\cdots,\phi_{D-m-1}),
\end{equation}
where, $f$ is an arbitrary function of the variables  $\phi_1,\phi_2,\cdots,\phi_{D-m-1}$ from the spherical coordinates
\begin{equation}\label{spherical}
\begin{aligned}
    &x_D=r\cos \phi_{D-1},\\
    &x_{D-1}=r \sin \phi_{D-1} \cos \phi_{D-2},\\
    &\cdots \cdots\\
    &x_2=r \sin \phi_{D-1} \sin \phi_{D-2} \cdots \sin \phi_{2} \cos \phi_{1},\\
    &x_1=r \sin \phi_{D-1} \sin \phi_{D-2} \cdots \sin \phi_{2} \sin \phi_{1}.
\end{aligned}
\end{equation}
So the extended hamiltonian (\ref{extended-H}) with $V$ satisfying the conditions (\ref{cd1}) and (\ref{cd2}) is given by
\begin{equation}\label{gernal}
    \hat{H}=-\sum_{i=1}^{D}\frac{\partial^2}{\partial x_i^2}-\frac{\eta}{r}+\frac{1}{\sum_{j=1}^{D-m}x_j^2}f(\phi_1,\phi_2,\cdots,\phi_{D-m-1}).
\end{equation}
This Hamiltonian possesses $m$ extended Laplace-Runge-Lenz vectors given by
\begin{equation}
\begin{aligned}
    X_i&=M_i-2x_iV\\
    &=\sum_{a=1}^{D}(p_aL_{ia}+L_{ia}p_a)+x_i\bigg(\frac{\eta}{r}-\frac{2}{\sum_{j=1}^{D-m}x_j^2}f(\phi_1,\phi_2,\cdots,\phi_{D-m-1}) \bigg),~~~i=D-m+1,\cdots,D.
\end{aligned}
\end{equation}
This is easily shown as follows. Using the conditions \eqref{cd1} and \eqref{cd2}, we have
\begin{equation}
    \begin{aligned}
    \bigg[X_i,\hat{H}\bigg]=&\bigg[\sum_{a=1}^{D}(p_aL_{ia}+L_{ia}p_a),V\bigg]+\bigg[-2x_iV,-\sum_{a=1}^{D}\frac{\partial^2}{\partial x_a^2} \bigg]\\
    =&\bigg[\sum_{a=1}^{D}\bigg\{2x_i\frac{\partial^2}{\partial x_a^2}-2x_a\frac{\partial}{\partial x_a}\frac{\partial}{\partial x_i}\bigg\},V \bigg]-2x_i\bigg[\sum_{a=1}^{D}\frac{\partial^2}{\partial x_a^2}, V \bigg]-4V\frac{\partial}{\partial x_i}\\
    =&2x_i\bigg[\sum_{a=1}^{D}\frac{\partial^2}{\partial x_a^2}, V \bigg]+4V\frac{\partial}{\partial x_i}-2x_i\bigg[\sum_{a=1}^{D}\frac{\partial^2}{\partial x_a^2}, V \bigg]-4V\frac{\partial}{\partial x_i}=0.
    \end{aligned}
\end{equation}
\vskip.1in

Thus we arrive at the main result of this section:
\begin{mypro}\label{p1}
A $D$-dimensional quantum system with Hamiltonian given by \eqref{gernal} admits $m$ extended and conserved Laplace-Runge-Lenz vectors 
\begin{equation}
\begin{aligned}
    &X_i=\sum_{a=1}^{D}(p_aL_{ia}+L_{ia}p_a)+x_i\bigg[\frac{\eta}{r}-\frac{2}{\sum_{j=1}^{D-m}x_j^2}f(\phi_1,\phi_2,\cdots,\phi_{D-m-1}) \bigg],\\
    &[X_i,\hat{H}]=0,~~~~~i=D-m+1,\cdots,D.
\end{aligned}
\end{equation}
\end{mypro}
When the function $f$ is equal to zero (for which the conditions \eqref{cd1} and \eqref{cd2} hold trivially), $X_i$'s reduce to the Laplace-Runge-Lenz vectors \eqref{glrl} for the $D$-dimensional hydrogen atom.

\section{New family of superintegrable systems in $D$ dimensions}
Let $\{x_1,x_2,\cdots,x_{D} \}$ denote the $D$ coordinates in the $D$-dimensional space. We divide the $D$ coordinates  into $N$ ($1 \leq N\leq D$) disjoint and nonempty blocks, say
\begin{equation}\label{partition}
    \begin{aligned}
    &\mathcal{B}_1=\{x_{n_0+1}=x_1,\cdots,x_{n_1} \},~~~n_0=0,\\
    &\mathcal{B}_2=\{x_{n_1+1},\cdots,x_{n_2} \},\\
    &\cdots\\
    &\mathcal{B}_{i+1}=\{x_{n_i+1},\cdots,x_{n_{i+1}} \},\\
    &\cdots\\
    &\mathcal{B}_{N-1}=\{x_{n_{N-2}+1},\cdots,x_{n_{N-1}} \},\\
    &\mathcal{B}_{N}=\{x_{n_{N-1}+1},\cdots,x_{n_N}=x_{D} \},~~~n_N=D
    \end{aligned}
\end{equation}
and throughout let 
\begin{equation}
d_i=|\mathcal{B}_i|=n_i-n_{i-1} \geq 1,~~~i=1,2,\cdots,N.
\end{equation}
We have $d_i \geq 1$ because $\mathcal{B}_i$ is nonempty.

We now propose the following new family of quantum systems
\begin{equation}\label{sys1}
\hat{H}=-\sum_{i=1}^{D}\frac{\partial^2}{\partial x_i^2}-\frac{\eta}{r}+\frac{\alpha_1}{x_1^2+\cdots+x_{n_1}^2}
+\frac{\alpha_2}{x_{n_1+1}^2+\cdots+x_{n_2}^2}+\cdots+\frac{\alpha_{N-1}}{x_{n_{N-2}+1}^2+\cdots+x_{n_{N-1}}^2},
\end{equation}
where $\alpha_1, \alpha_2, \cdots, \alpha_{N-1}$ are real parameters.  
Introducing the notation
\begin{equation}\label{ri}
    r_i=\sqrt{\sum_{x\in \mathcal{B}_i}x^2},~~~i=1,2,\cdots,N,
\end{equation}
then (\ref{sys1}) can be expressed in the more compact form
\begin{equation}\label{sys}
    \hat{H}=-\sum_{i=1}^{D}\frac{\partial^2}{\partial x_i^2}-\frac{\eta}{r}+\sum_{i=1}^{N-1}\frac{\alpha_i}{r_i^2},~~~.
\end{equation}
Note that in terms of $r_i$, we have $r=\sqrt{\sum_{i=1}^{D}x_i^2}=\sqrt{\sum_{i=1}^{N}r_i^2}$.
When $N=1$ the above system reduces to the hydrogen atom (\ref{atom}), while when $N=D$, i.e. each block contains exactly one coordinate, we recover the $D$-dimensional system introduced in \cite{RW}
\begin{equation}\label{N=D}
     {\cal H}=-\sum_{i=1}^{D}\frac{\partial^2}{\partial x_i^2}-\frac{\eta}{r}+\sum_{i=1}^{D-1}\frac{\alpha_i}{x_i^2}.
\end{equation}
So our system (\ref{sys}) contain both the hydrongen atom and the model in \cite{RW} as special cases.

The first set of integrals of the system (\ref{sys}) consists of the angular momenta from each block $\mathcal{B}_k$
\begin{equation}\label{i1}
\begin{aligned}
    &L_{ij}=x_ip_j-x_jp_i,~~~p_i=\frac{\partial}{\partial x_i},~~~p_j=\frac{\partial}{\partial x_j}~~~x_i,x_j\in \mathcal{B}_k,~~~k=1,\cdots,N.\\
\end{aligned}
\end{equation}
Notice that $V=\sum_{i=1}^{N-1}\alpha_i/r_i^2$ satisfies the conditions \eqref{cd1} and \eqref{cd2} for $m=d_N$ and takes the form of (\ref{V}). This is seen as follows.
\begin{equation}
 \sum_{i=1}^{N-1}\alpha_i/r_i^2=\frac{1}{\sum_{j=1}^{D-d_N}x_j^2}\sum_{i=1}^{N-1}\alpha_i/r_i^2 \times 
 \sum_{j=1}^{D-d_N}x_j^2=\frac{1}{\sum_{j=1}^{D-d_N}x_j^2}\sum_{i=1}^{N-1}\sum_{j=1}^{n_{N-1}}\frac{\alpha_i x_j^2}    {r_i^2}.
\end{equation} 
In terms of the coordinates (\ref{spherical}), it is then obvious from (\ref{partition}) and (\ref{ri}) that the double summation on the right hand side of the above equation is some function of angles $\phi_1, \phi_2, \cdots, \phi_{D-d_N-1}$ only.
So our Hamiltonian  \eqref{sys} has the form \eqref{gernal}, and we can apply the proposition \eqref{p1} to obtain $d_N$ extended and conserved Laplace-Runge-Lenz vectors of the system
\begin{equation}\label{i2}
    \hat{X}_j=\sum_{a=1}^{D}(p_aL_{ja}+L_{ja}p_a)+x_j\bigg(\frac{\eta}{r}-\sum_{i=1}^{N-1}\frac{2\alpha_i}{r_i^2} \bigg),~~~j=D-d_N+1,\cdots,D.
\end{equation}
For convenience, we also define
\begin{equation}
    \hat{X}_j=0,~~~j=1,2,\cdots,D-d_N.
\end{equation}
It can be checked that the system (\ref{sys}) has also the following two sets of integrals of motion
\begin{equation}\label{i3}
\begin{aligned}
    &\hat{Z}_l=\sum_{1\leq i<k\leq n_l}L_{ik}^2-\bigg(\sum_{i=1}^{l}r_i^2\bigg)\bigg(\sum_{i=1}^{l}\frac{\alpha_i}{r_i^2} \bigg),~~~l=2,\cdots,N-1,\\
\end{aligned}
\end{equation}
\begin{equation}\label{i4}
    \hat{Y}_p=\sum_{n_{p-1}+1\leq i<k\leq D}L_{ik}^2-\bigg(\sum_{i=p}^{N}r_i^2 \bigg)\bigg(\sum_{i=p}^{N-1}\frac{\alpha_i}{r_i^2} \bigg),~~~p=1,\cdots,N-1,
\end{equation}
where the numbers $n_l$ and $n_{p-1}+1$ are the largest and smallest indices in $\mathcal{B}_l$ and $\mathcal{B}_p$, respectively. Together with $\hat{H}$, \eqref{i1}, \eqref{i2}, \eqref{i3} and \eqref{i4} give $D+N+d_N-2$ algebraically independent integrals of motion in total, and thus the system (\ref{sys}) is superintegrable. 

Some remarks are in order. When $N=D$, only one extended and conserved Laplace-Runge-Lenz vector survives, and (\ref{i2}), (\ref{i3}) and (\ref{i4}) reduce to the integrals obtained in \cite{RW} for the superintegrable system (\ref{N=D}),
\begin{equation}\label{wlrl}
    X=\sum_{k=1}^{D}(p_kL_{Dk}+L_{Dk}p_k)+x_D\bigg(\frac{\eta}{r}-\sum_{i=1}^{D-1}\frac{2\alpha_i}{x_i^2}\bigg),
\end{equation}
\begin{equation}\label{oldz}
    Z_l=\sum_{1\leq i<k\leq l}L^2_{ik}-\bigg(\sum_{i=1}^{l}x_i^2\bigg)\bigg(\sum_{k=1}^{l}\frac{\alpha_i}{x_i^2}\bigg),~~~l=2,\cdots,D-1,
\end{equation}
\begin{equation}\label{oldy}
    Y_p=\sum_{p\leq i<k \leq D}L^2_{ik}-\bigg(\sum_{i=p}^{D}x_i^2 \bigg)\bigg(\sum_{k=p}^{D-1} \frac{\alpha_i}{x_k^2}\bigg),~~~p=1,2,\cdots,D-1,
\end{equation}
In this case the total number of independent integrals of motion becomes $2D-1$.



\section{Separation of variables and energy spectrum}

In order to solve the eigenvalue problem $\hat{H}\Psi=E\Psi$, we use the polar coordinates in each $\mathcal{B}_i$,
$i=1,2,\cdots, N$, 
\begin{equation}\label{polar1}
\begin{aligned}
    &x_{n_i}=r_i\cos \phi^{(i)}_{d_i-1},\\
    &x_{n_i-1}=r_i\sin \phi^{(i)}_{d_i-1}\cos \phi^{(i)}_{d_i-2},\\
    &\cdots \cdots\\
    &x_{n_{i-1}+1}=r_i\sin \phi^{(i)}_{d_i-1} \sin \phi^{(i)}_{d_i-2} \cdots \sin \phi^{(i)}_1.
\end{aligned}
\end{equation}
In these coordinates, we have
\begin{equation}
   -\sum_{x\in \mathcal{B}_i}\frac{\partial^2}{\partial x^2}=-\frac{\partial^2}{\partial r_i^2}-\frac{d_i-1}{r_i}\frac{\partial}{\partial r_i}-\frac{1}{r_i^2}\hat{L}_i^2,~~~i=1,2,\cdots,N,
\end{equation}
where $\hat{L}_i^2$ is the total angular momenta associated with the block $\mathcal{B}_i$
\begin{equation}
   \hat{L}_i^2=\sum_{k<l}\bigg(x_k\frac{\partial}{\partial x_l}-x_l\frac{\partial}{x_k} \bigg)^2,~~~x_k,x_l \in \mathcal{B}_i.
\end{equation}
The Hamiltonian \eqref{sys} is transformed to
\begin{equation}\label{pol}
    \hat{H}=\sum_{i=1}^{N}\bigg\{-\frac{\partial^2}{\partial r_i^2}-\frac{d_i-1}{r_i}\frac{\partial}{\partial r_i}-\frac{1}{r_i^2}\hat{L}_i^2 \bigg\}-\frac{\eta}{r}+\sum_{i=1}^{N-1}\frac{\alpha_i}{r_i^2}.
\end{equation}
We write
\begin{equation}
    \Psi=R(r_1,\cdots,r_{N})\prod_{i=1}^{N}r_i^{-\frac{d_i-1}{2}}\prod_{i=1}^{N}Y^{(i)},
\end{equation}
where $Y^{(i)}$ are the spherical harmonics in $d_i$-dimensional space, which satisfy 
\begin{equation}
\begin{aligned}
    (-\hat{L}_i^2+\alpha_i)Y^{(i)}=&[l_i(l_i+d_i-2)+\alpha_i]Y^{(i)},\\
    &i=1,2,\cdots,N,~~~~l_i=0,1,2,\cdots,~~~~\alpha_N=0.
\end{aligned}
\end{equation}
$R(r_1,\cdots,r_{N})$ is determined by
\begin{equation}\label{R}
\begin{aligned}
    &\bigg[-\sum_{i=1}^{N}\frac{\partial^2}{\partial r_i^2}-\frac{\eta}{r}+\sum_{i=1}^{N}\frac{\lambda_i}{r_i^2}\bigg]R(r_1,\cdots,r_{N})=ER(r_1,\cdots,r_{N}),\\
    &~~~~~~~~~~~~~~~~~~~~\lambda_i=l_i(l_i+d_i-2)+\alpha_i+\frac{1}{4}(d_i-1)(d_i-3).
\end{aligned}
\end{equation}
To solve \eqref{R}, we regard it as a $N$-dimensional system and use the polar coordinates
\begin{equation}
\begin{aligned}
    &r_N=r\cos \theta_{N-1},\\
    &r_{N-1}=r \sin \theta_{N-1} \cos \theta_{N-2},\\
    &\cdots \cdots\\
    &r_1=r \sin \theta_{N-1} \sin \theta_{N-2} \cdots \sin \theta_1.
\end{aligned}
\end{equation}
Above coordinates allow us to factorize $R(r_1,\cdots,r_{N})$
\begin{equation}
    R(r_1,\cdots,r_{N})=r^{-\frac{N-1}{2}}F(r)\prod_{i=1}^{N-1}y_i(\theta_i),
\end{equation}
where $y_i(\theta_i)$ and $F(r)$ are determined by the differential equations
\begin{equation}
    \begin{aligned}
    &\bigg[-\frac{\partial^2}{\partial \theta_{N-1}^2}-(N-2)\frac{\cos \theta_{N-1}}{\sin \theta_{N-1}}\frac{\partial}{\partial \theta_{N-1}} +\frac{b_{N-2}}{\sin^2 \theta_{N-1}}+\frac{\lambda_{N}}{\cos^2 \theta_{N-1}}\bigg]y_{N-1}(\theta_{N-1})=b_{N-1}y_{N-1}(\theta_{N-1}),\\
    &\bigg[-\frac{\partial^2}{\partial \theta_{N-2}^2}-(N-3)\frac{\cos \theta_{N-2}}{\sin \theta_{N-2}}\frac{\partial}{\partial \theta_{N-2}} +\frac{b_{N-3}}{\sin^2 \theta_{N-2}}+\frac{\lambda_{N-1}}{\cos^2 \theta_{N-2}}\bigg]y_{N-2}(\theta_{N-2})=b_{N-2}y_{N-2}(\theta_{N-2}),\\
    &\cdots \cdots\\
    &\bigg[-\frac{\partial^2}{\partial \theta_{2}^2}-\frac{\cos \theta_{2}}{\sin \theta_{2}}\frac{\partial}{\partial \theta_{2}} +\frac{b_1}{\sin^2 \theta_2}+\frac{\lambda_3}{\cos^2 \theta_2}\bigg]y_2(\theta_2)=b_2y_2(\theta_2),\\
    &\bigg[-\frac{\partial^2}{\partial \theta_1^2}+\frac{\lambda_1}{\sin^2 \theta_1}+\frac{\lambda_2}{\cos^2 \theta_1}\bigg]y_1(\theta_1)=b_1y_1(\theta_1)
    \end{aligned}
\end{equation}
and
\begin{equation}
    \bigg[-\frac{\partial^2}{\partial r^2}-\frac{\eta}{r}+\frac{b_{N-1}+(N-1)(N-3)/4}{r^2}\bigg]F(r)=EF(r)
\end{equation}
The equations for the angles can be solved in terms of Jacobi polynomials
\begin{equation}
y_i(\theta_i)=(\sin \theta_i)^{\kappa_{i-1}+1-\frac{i}{2}} (\cos \theta_i)^{\gamma_{i+1}+\frac{1}{2}}P^{(\kappa_{i-1},\gamma_{i+1})}_{J_i}(\cos 2\theta_i),~~~~~i=1,2,\cdots,N-1,
\end{equation}
with $b_i$ given by $b_i=\kappa_i^2-\frac{(i-1)^2}{4}$, where $J_i=0,1,\cdots$, and 
\begin{equation}
   \kappa_{i}=2\sum_{s=1}^{i}J_s+i+\sum_{s=1}^{i+1}\gamma_s,~~~~~\gamma_s=\frac{\sqrt{1+4\lambda_s}}{2}.
\end{equation}
Solution to the radial part $F(r)$ is given by
\begin{equation}
    \begin{aligned}
    &F(r)=r^{\kappa}e^{-\sqrt{-E}r}L^{(2\kappa-1)}_{N_r}(2\sqrt{-E}r),~~~~\kappa=2\sum_{s=1}^{N-1}J_s+N-\frac{1}{2}+\frac{1}{2}\sum_{s=1}^{N}\sqrt{1+4\lambda_s},\\
    \end{aligned}
\end{equation}
where $L^{(2\kappa-1)}_{N_r}$ is the Laguerre polynomial. The final expression for the eigenfuntion $\Psi$ is
\begin{equation}
\begin{aligned}
    \Psi_{N_r,J_1,\cdots,J_{N-1},l_1,\cdots,l_{N}}=& ~ r^{\kappa-\frac{N-1}{2}}e^{-\sqrt{-E}r}L^{(2\kappa-1)}_{N_r}(2\sqrt{-E}r)\times \\
    &\prod_{i=1}^{N-1}(\sin \theta_i)^{\kappa_{i-1}+1-\frac{i}{2}} (\cos \theta_i)^{\gamma_{i+1}+\frac{1}{2}}P^{(\kappa_{i-1},\gamma_{i+1})}_{J_i}(\cos 2\theta_i)\times \prod_{i=1}^{N}r_i^{-\frac{d_i-1}{2}}\times \prod_{i=1}^{N}Y^{(i)}
\end{aligned}
\end{equation}
The corresponding energy spectrum is given by the shifted Balmer formula
\begin{equation}
    E=-\frac{\eta^2}{(2N_r+4\sum_{s=1}^{N-1}J_s+2N-1+2\sum_{s=1}^{N}\gamma_s)^2},~~~~N_r=0,1,\cdots.
\end{equation}


\section{Quadratic algebra structure}

It can be shown that the integrals (\ref{i2}), (\ref{i3}) and (\ref{i4}) satisfy following commutation relations
\begin{equation}
    \begin{aligned}
    &\comm{\hat{Y}_i}{\hat{Y}_j}=0=\comm{\hat{X}_i}{\hat{Z}_j},~~~\comm{\hat{Y}_1}{\hat{Z}_i}=0=\comm{\hat{Z}_i}{\hat{Z}_j},\\
    \end{aligned}
\end{equation}
The partition of the coordinates implies the system (\ref{sys}) admits Lie algebra $so(d_1)\oplus so(d_2)\cdots \oplus so(d_N) $ symmetry, where $so(d_p)$ is generated by angular momenta $L_{ij}$ (\ref{i1}) associated with the block $\mathcal{B}_p$,
\begin{equation}
    \comm{L_{ij}}{L_{kl}}=\delta_{jk}L_{il}+\delta_{il}L_{jk}-\delta_{ik}L_{jl}-\delta_{jl}L_{ik}.
\end{equation}
In addition, for the conserved angular momenta $L_{ij}$ and $\hat{X}_k$'s, we have
\begin{equation}
\begin{aligned}
    &\comm{L_{ij}}{\hat{X}_k}=\delta_{jk}\hat{X}_i-\delta_{ik}\hat{X}_j,\\
    &\comm{\hat{X}_i}{\hat{X}_j}=-4\hat{H}L_{ij}.
\end{aligned}
\end{equation}
In what follows, we present the quadratic algebra relations among the integrals.
Recall that the total angular momenta for block $\mathcal{B}_p$ is
\begin{equation}
    \hat{L}_p^2=\sum_{i<j}\bigg(x_i\frac{\partial}{\partial x_j}-x_j\frac{\partial}{\partial x_i}\bigg)^2,~~~x_i,x_j \in \mathcal{B}_p.
\end{equation}
We define the number $\mathcal{N}_p$ 
\begin{equation}
    \mathcal{N}_p=\bigg(\sum_{i=1}^{p}d_i-2\bigg)\bigg(\sum_{i=1}^{p}\frac{d_i-1}{2} \bigg)-\bigg(\sum_{i=1}^{p}\frac{d_i-1}{2} \bigg)^2,~~p=2,\cdots,N-1
\end{equation}
and the number $\mathcal{M}_p$ 
\begin{equation}
\begin{aligned}
    &\mathcal{M}_p=\bigg(\sum_{i=p}^{N}d_i-2\bigg)\bigg(\sum_{i=p}^{N-1}\frac{d_i-1}{2} \bigg)-\bigg(\sum_{i=p}^{N-1}\frac{d_i-1}{2} \bigg)^2,~~~p=1,\cdots,N-1,\\
\end{aligned}
\end{equation}

We now state the main result in this section.

\begin{mypro}\label{p2}
For $j=D-d_N+1,\cdots,D$, $\hat{Y}_1$ and $\hat{X}_j$ satisfy the commutation relations
\begin{equation}\label{qa1}
\begin{aligned}
    \comm{\hat{Y}_1}{\hat{X}_j}=&\hat{W}_j,\\
    \comm{\hat{Y}_1}{\hat{W}_j}=&-2\acomm{\hat{Y}_1}{\hat{X}_j}+(D-3)(D-1)\hat{X}_j,\\
    \comm{\hat{X}_j}{\hat{W}_j}=&2\hat{X}_j^2-8\hat{H}(\hat{Z}_{N-1}-\mathcal{N}_{N-1})+16(\hat{Y}_1-\mathcal{M}_1)\hat{H}-2(N+d_N-2)^2\hat{H}-2\eta^2.
\end{aligned}
\end{equation}
For $2\leq p \leq N-1$, the integrals $\hat{Y}_p$ and $\hat{Z}_p$ overlap in block $\mathcal{B}_p$. They obey the quadratic algebra relations 
\begin{equation}\label{qa2}
\begin{aligned}
   \comm{\hat{Z}_p}{\hat{Y}_p}=&\hat{C}_p,\\ 
    \comm{\hat{Z}_p}{\hat{C}_p}=&-8(\hat{Z}_p-\mathcal{N}_p)^2-8 \acomm{\hat{Z}_p-\mathcal{N}_p}{\hat{Y}_p-\mathcal{M}_p}\\
    &-4\bigg[(p-2)(N+d_N-1)-p^2+p+4+2\bigg(-\hat{L}_p^2+\alpha_p+\frac{1}{4}(d_p-1)(d_p-3) \bigg) \bigg](\hat{Z}_p-\mathcal{N}_p)\\
    &+4(N+d_N-p)(N+d_N-p-4)(\hat{Y}_p-\mathcal{M}_p)+8(\hat{Y}_1-\mathcal{M}_1+\hat{Z}_{p-1}-\mathcal{N}_{p-1})(\hat{Z}_p-\mathcal{N}_p)\\
    &-4\bigg[N+d_N-p-4+2\bigg(-\hat{L}_p^2+\alpha_p+\frac{1}{4}(d_p-1)(d_p-3) \bigg) \bigg](\hat{Y}_1-\mathcal{M}_1)\\
    &-4\bigg[(N+d_N-p-1)(N+d_N-p-4)-2\bigg(-\hat{L}_p^2+\alpha_p+\frac{1}{4}(d_p-1)(d_p-3) \bigg) \bigg](\hat{Z}_{p-1}-\mathcal{N}_{p-1})\\
    &+4(N+d_N-p)(N+d_N-5)\bigg(-\hat{L}_p^2+\alpha_p+\frac{1}{4}(d_p-1)(d_p-3) \bigg)\\
    &+4(p-1)(N+d_N-p)(\hat{Y}_{p+1}-\mathcal{M}_{p+1})-8(\hat{Y}_1-\mathcal{M}_1)(\hat{Y}_{p+1}-\mathcal{M}_{p+1})\\
    &+8(\hat{Z}_p-\mathcal{N}_p)(\hat{Y}_{p+1}-\mathcal{M}_{p+1})+8(\hat{Z}_{p-1}-\mathcal{N}_{p-1})(\hat{Y}_{p+1}-\mathcal{M}_{p+1}),
\end{aligned}
\end{equation}
\begin{equation}\label{qa3}
\begin{aligned}
    \comm{\hat{Y}_p}{\hat{C}_p}=&+8(\hat{Y}_p-\mathcal{M}_p)^2+8\acomm{\hat{Z}_p-\mathcal{N}_p}{\hat{Y}_p-\mathcal{M}_p}-4p(p-4)(\hat{Z}_p-\mathcal{N}_p)\\
    &+4\bigg[(p-2)(N+d_N-1)-p^2+p+4+2\bigg(-\hat{L}_p^2+\alpha_p+\frac{1}{4}(d_p-1)(d_p-3) \bigg) \bigg](\hat{Y}_1-\mathcal{M}_p)\\
    &-8(\hat{Z}_{p-1}-\mathcal{N}_{p-1})(\hat{Y}_p-\mathcal{M}_p)-8(\hat{Y}_{1}-\mathcal{M}_1)(\hat{Y}_p-\mathcal{M}_p)\\
    &+4\bigg[p-4+2\bigg(-\hat{L}_p^2+\alpha_p+\frac{1}{4}(d_p-1)(d_p-3) \bigg) \bigg](\hat{Y}_1-\mathcal{M}_1)+8(\hat{Z}_{p-1}-\mathcal{N}_{p-1})(\hat{Y}_1-\mathcal{M}_1)\\
    &-4p(N+d_N-p-1)(\hat{Z}_{p-1}-\mathcal{N}_{p-1})-4p(N+d_N-5)\bigg(-\hat{L}_p^2+\alpha_p+\frac{1}{4}(d_p-1)(d_p-3) \bigg)\\
    &+4\bigg[(p-4)(p-1)-2\bigg(-\hat{L}_p^2+\alpha_p+\frac{1}{4}(d_p-1)(d_p-3) \bigg) \bigg](\hat{Y}_{p+1}-\mathcal{M}_{p+1})\\
    &-8(\hat{Z}_{p-1}-\mathcal{N}_{p-1})(\hat{Y}_{p+1}-\mathcal{M}_{p+1})-8(\hat{Y}_{p+1}-\mathcal{M}_{p+1})(\hat{Y}_p-\mathcal{M}_p).
\end{aligned}
\end{equation}
It should be understood that $Z_1=-\alpha_1$ and $Y_N=0$.
\end{mypro}

In the remaining space of this section we outline the steps of proof of this proposition. Let us first of all recall that when $N=D$, the integrals (\ref{i2}), (\ref{i3}) and (\ref{i4}) reduce to (\ref{wlrl}), (\ref{oldz}) and (\ref{oldy}) for the  system (\ref{N=D}) in \cite{RW}. 
As shown in \cite{LMZ}, these later integrals satisfy the quadratic algebra relations:
\begin{equation}
    \begin{aligned}
    \comm{Y_1}{X}=&C_1,\\
    \comm{Y_1}{C_1}=&-2\acomm{Y_1}{X}+(D-3)(D-1)X,\\
    \comm{X}{C_1}=&2X^2-8{\cal H}Z_{D-1}+16Y_1{\cal H}-(D-1)^2{\cal H}-2\eta^2,
    \end{aligned}
\end{equation}
and for each pair $(Z_p, Y_p)$ with $2\leq p\leq D-1$,
\begin{equation}\label{oldq1}
\begin{aligned}
    \comm{Z_p}{Y_p}=&C_p,\\
    \comm{Z_p}{C_p}=&-8Z_p^2-8\acomm{Z_p}{Y_p}-4((p-2)D-p^2+p+4+2\alpha_p)Z_p\\
    &+4(D-p+1)(D-p-3)Y_p+8(Y_1+Z_{p-1})Z_p-4(D-p-3+2\alpha_p)Y_1\\
    &-4((D-p)(D-p-3)-2\alpha_p)Z_{p-1}+4(D-p+1)(D-4)\alpha_p\\
    &+4(p-1)(D-p+1)Y_{p+1}-8Y_1Y_{p+1}+8Z_pY_{p+1}+8Z_{p-1}Y_{p+1},
\end{aligned}
\end{equation}
\begin{equation}\label{oldqp}
\begin{aligned}    
    \comm{Y_p}{C_p}=&+8Y_p^2+8\acomm{Z_p}{Y_p}-4p(p-4)Z_p+4((p-2)D-p^2+p+4+2\alpha_p)Y_p\\
    &-8Z_{p-1}Y_p-8Y_1Y_p+4(p-4+2\alpha_p)Y_1+8Z_{p-1}Y_1-4p(D-p)Z_{p-1}\\
    &-4p(D-4)\alpha_p+4((p-4)(p-1)-2\alpha_p)Y_{p+1}-8Z_{p-1}Y_{p+1}-8Y_{p+1}Y_p.
\end{aligned}
\end{equation}
In the above relations, it is understood that $Z_1=-\alpha_1$ and $Y_D=0$.

\vskip.1in
\noindent\underline{\textbf{Step 1:}}
We introduce the operators $\mathcal{S}_{i,j}$ for integers $j>i$ and parameters $q_i,\cdots, q_j$,
\begin{equation}
\begin{aligned}
    \mathcal{S}_{i,j}(q_i,\cdots,q_{j})=&(x_i^2+\cdots+x_j^2)\bigg(\frac{\partial^2}{\partial x_i^2}+\cdots+\frac{\partial^2}{\partial x_j^2}\bigg)-\bigg(x_i\frac{\partial}{\partial x_i}+\cdots+x_j\frac{\partial}{\partial x_j}\bigg)^2\\
    &-(j-i-1)\bigg(x_i\frac{\partial}{\partial x_i}+\cdots+x_j\frac{\partial}{\partial x_j}\bigg)
    -(x_i^2+\cdots+x_j^2)\bigg(\frac{q_i}{x_i^2}+\cdots+\frac{q_j}{x_j^2} \bigg),
\end{aligned}
\end{equation}
and expand \eqref{oldz} and \eqref{oldy} in the form
\begin{equation}
\begin{aligned}
    Z_p=&(x_1^2+\cdots x_p^2)\bigg(\frac{\partial^2}{\partial x_1^2}+\cdots+\frac{\partial^2}{\partial x_p^2}\bigg)-\bigg(x_1\frac{\partial}{\partial x_1}+\cdots+x_p\frac{\partial}{\partial x_p}\bigg)^2\\
    &-(p-2)\bigg(x_1\frac{\partial}{\partial x_1}+\cdots+x_p\frac{\partial}{\partial x_p}\bigg)-(x_1^2+\cdots+x_p^2)\bigg(\frac{\alpha_1}{x_1^2}+\cdots+\frac{\alpha_p}{x_p^2} \bigg),\\
    Y_p=&(x_p^2+\cdots x_D^2)\bigg(\frac{\partial^2}{\partial x_p^2}+\cdots+\frac{\partial^2}{\partial x_D^2}\bigg)-\bigg(x_p\frac{\partial}{\partial x_p}+\cdots+x_D\frac{\partial}{\partial x_D}\bigg)^2\\
    &-(D-p-1)\bigg(x_p\frac{\partial}{\partial x_p}+\cdots+x_D\frac{\partial}{\partial x_D}\bigg)-(x_p^2+\cdots+x_D^2)\bigg(\frac{\alpha_p}{x_p^2}+\cdots+\frac{\alpha_{D-1}}{x_{D-1}^2}+\frac{0}{x_D^2} \bigg).\\
\end{aligned}
\end{equation}
In terms of ${\cal S}_{ij}$, the integrals $Z_p$ and $Y_p$ can be expressed as 
\begin{equation}
    Z_p=\mathcal{S}_{1,p}(\alpha_1,\cdots,\alpha_p),~~~Y_p=\mathcal{S}_{p,D}(\alpha_p,\cdots,\alpha_D=0),
\end{equation}
where $2\leq p\leq D-1$. Then the commutation relations \eqref{oldq1} and \eqref{oldqp} can be written as
\begin{equation}\label{qform}
\begin{aligned}
    &\comm{\mathcal{S}_{1,p}}{\comm{\mathcal{S}_{1,p}}{\mathcal{S}_{p,D}}}=\mathcal{Q}_1(p,D,\alpha_p,\mathcal{S}_{1,D},\mathcal{S}_{p,D},\mathcal{S}_{p+1,D},\mathcal{S}_{1,p-1},\mathcal{S}_{1,p}),\\
    &\comm{\mathcal{S}_{p,D}}{\comm{\mathcal{S}_{1,p}}{\mathcal{S}_{p,D}}}=\mathcal{Q}_2(p,D,\alpha_p,\mathcal{S}_{1,D},\mathcal{S}_{p,D},\mathcal{S}_{p+1,D},\mathcal{S}_{1,p-1},\mathcal{S}_{1,p}),\\
\end{aligned}
\end{equation}
where $\mathcal{Q}_1$ and $\mathcal{Q}_2$ represent the expressions on the right hand side of \eqref{oldq1} and \eqref{oldqp}, respectively.

\vskip.1in
\noindent \underline{\textbf{Step 2:}} We use the polar coordinates in $\mathcal{B}_1,\cdots,\mathcal{B}_{N-1}$ to transform the integrals (\ref{i3}) and (\ref{i4}) into 
\begin{equation}\label{61a}
\begin{aligned}
    \hat{Z}_p=&~ \bigg(\sum_{i=1}^{p}r_i^2\bigg)\bigg\{\sum_{i=1}^{p}\bigg(\frac{\partial^2}{\partial r_i^2}+\frac{d_i-1}{r_i}\frac{\partial}{\partial r_i}+\frac{1}{r_i^2}\hat{L}_i^2 \bigg) \bigg\}-\bigg(\sum_{i=1}^{p}r_i^2\bigg)\bigg(\frac{\alpha_1}{r_1^2}+\cdots+\frac{\alpha_{p}}{r_{p}^2}\bigg)\\
    &-\bigg(\sum_{i=1}^{p}r_i\frac{\partial}{\partial r_i}\bigg)^2-\bigg(\sum_{i=1}^{p}d_i-2\bigg)\bigg(\sum_{i=1}^{p}r_i\frac{\partial}{\partial r_i}\bigg),
\end{aligned}
\end{equation}
\begin{equation}\label{61}
\begin{aligned}
    \hat{Y}_p=&~\bigg(\sum_{i=p}^{N-1}r_i^2+x_{D-d_N+1}^2+\cdots+x_{D}^2 \bigg)\\
  &~~\times\bigg\{\sum_{i=p}^{N-1}\bigg(\frac{\partial^2}{\partial r_i^2}+\frac{d_i-1}{r_i}\frac{\partial}{\partial r_i}+\frac{1}{r_i^2}\hat{L}_i^2 \bigg)+\frac{\partial^2}{\partial x_{D-d_N+1}^2}+\cdots+\frac{\partial^2}{\partial x_{D}^2} \bigg\}\\
    &-\bigg(\sum_{i=p}^{N-1}r_i\frac{\partial}{\partial r_i}+x_{D-d_N+1}\frac{\partial}{\partial x_{D-d_N+1}}+\cdots+x_{D}\frac{\partial}{x_{D}}\bigg)^2\\
    &-\bigg(\sum_{i=p}^{N}d_i-2 \bigg)\bigg(\sum_{i=p}^{N-1}r_i\frac{\partial}{\partial r_i}+x_{D-d_N+1}\frac{\partial}{\partial x_{D-d_N+1}}+\cdots+x_{D}\frac{\partial}{\partial x_{D}}\bigg)\\
    &-\bigg(\sum_{i=p}^{N-1}r_i^2+x_{D-d_N+1}^2+\cdots+x_{D}^2 \bigg)\bigg(\frac{\alpha_p}{r_p^2}+\cdots+\frac{\alpha_{N-1}}{r_{N-1}^2}+\frac{0}{x_{D-d_N+1}^2}+\cdots+\frac{0}{x_{D-1}^2} \bigg).
\end{aligned}
\end{equation}
\\
\underline{\textbf{Step 3:}} Using similarity transformations to cancel the terms of first order derivatives, we have
\begin{equation}
\begin{aligned}
    \prod_{i=1}^{N-1}r_i^{(d_i-1)/2}\hat{Z}_p\prod_{i=1}^{N-1}r_i^{-(d_i-1)/2}=&~\bigg(\sum_{i=1}^{p}r_i^2 \bigg)\bigg(\sum_{i=1}^{p}\frac{\partial^2}{\partial r_i^2}  \bigg)-\bigg(\sum_{i=1}^{p}r_i\frac{\partial}{\partial r_i}\bigg)^2-(p-2)\bigg(\sum_{i=1}^{p}r_i\frac{\partial}{\partial r_i}\bigg)\\
    &-\bigg(\sum_{i=1}^{p}r_i^2\bigg)\bigg(\frac{c_1}{r_1^2}+\cdots+\frac{c_{p}}{r_{p}^2}\bigg)+\mathcal{N}_p=\mathcal{S}_{1,p}(c_1,\cdots,c_p)+\mathcal{N}_p, 
\end{aligned}
\end{equation}
where 
$$c_1=-\hat{L}_1^2+\alpha_1+\frac{1}{4}(d_1-1)(d_1-3),~~\cdots \cdots~~c_p=-\hat{L}_p^2+\alpha_p+\frac{1}{4}(d_p-1)(d_p-3)$$
Similarly,
\begin{equation}\label{gauge-Y}
    \begin{aligned}
     \prod_{i=1}^{N-1}r_i^{(d_i-1)/2}\hat{Y}_p\prod_{i=1}^{N-1}r_i^{-(d_i-1)/2}=&~\bigg(\sum_{i=p}^{N-1}r_i^2+x_{D-d_N+1}^2+\cdots+x_{D}^2 \bigg)\\
     &~~\times \bigg\{\sum_{i=p}^{N-1}\bigg(\frac{\partial^2}{\partial r_i^2} \bigg)+\frac{\partial^2}{\partial x_{D-d_N+1}^2}+\cdots+\frac{\partial^2}{\partial x_{D}^2} \bigg\}\\
    &-\bigg(\sum_{i=p}^{N-1}r_i\frac{\partial}{\partial r_i}+x_{D-d_N+1}\frac{\partial}{\partial x_{D-d_N+1}}+\cdots+x_{D}\frac{\partial}{x_{D}}\bigg)^2\\
    &-(N+d_N-3 )\bigg(\sum_{i=p}^{N-1}r_i\frac{\partial}{\partial r_i}+x_{D-d_N+1}\frac{\partial}{\partial x_{D-d_N+1}}+\cdots+x_{D}\frac{\partial}{\partial x_{D}}\bigg)\\
    &-\bigg(\sum_{i=p}^{N-1}r_i^2+x_{D-d_N+1}^2+\cdots+x_{D}^2 \bigg)\\
    &~~\times\bigg(\frac{c_p}{r_p^2}+\cdots+\frac{c_{N-1}}{r_{N-1}^2}+\frac{c_N}{x_{D-d_N+1}^2}+\cdots+\frac{c_{N+d_N-2}}{x_{D-1}^2}+\frac{0}{x_D^2}\bigg)+\mathcal{M}_p   
     \end{aligned}
\end{equation}  
where $c_N=c_{N+1}=\cdots =c_{N+d_N-1}\equiv 0$ and
\begin{equation}
    c_p=-\hat{L}_p^2+\alpha_p+\frac{1}{4}(d_p-1)(d_p-3),~~\cdots,~~
     c_{N-1}=-\hat{L}_{N-1}^2+\alpha_{N-1}+\frac{1}{4}(d_{N-1}-1)(d_{N-1}-3),
\end{equation}
which can be regarded as constants since they are central elements. Moreover we can conveniently treat the variables $x_{D-d_N+1}$, $x_{D-d_N+2}$, $\cdots$, $x_{D}$  in (\ref{gauge-Y})
 as $r_N$,  $r_{N+1}$, $\cdots$, $r_{N+d_N-1}$, respectively. Then  
\begin{equation}
 {\rm the~ right~ hand~ side~ of~(\ref{gauge-Y})} =\mathcal{S}_{p,N+d_N-1}(c_p,\cdots,c_{N+d_N-1}=0)+\mathcal{M}_p.
\end{equation}
\\
\underline{\textbf{Step 4:}} Substituting into \eqref{qform}, it is not hard to conclude
\begin{equation}
\begin{aligned}
    &\comm{\hat{Z}_p}{\comm{\hat{Z}_p}{\hat{Y}_p}}=\mathcal{Q}_1(p,N+d_N-1,c_p,\hat{Y}_1-\mathcal{M}_1,\hat{Y}_p-\mathcal{M}_p,\hat{Y}_{p+1}-\mathcal{M}_{p+1},\hat{Z}_{p-1}-\mathcal{N}_{p-1},\hat{Z}_p-\mathcal{N}_p),\\
    &\comm{\hat{Y}_p}{\comm{\hat{Z}_p}{\hat{Y}_p}}=\mathcal{Q}_2(p,N+d_N-1,c_p,\hat{Y}_1-\mathcal{M}_1,\hat{Y}_p-\mathcal{M}_p,\hat{Y}_{p+1}-\mathcal{M}_{p+1},\hat{Z}_{p-1}-\mathcal{N}_{p-1},\hat{Z}_p-\mathcal{N}_p),
\end{aligned}
\end{equation}
which are nothing but the commutation relations (\ref{qa2}) and (\ref{qa3}), respectively. 
The quadratic algebra relations (\ref{qa1}) generated by $\hat{Y}_1$ and $\hat{X}_j$ can be derived in a similar way. This completes our proof to the proposition.


\section{Conclusions}
We have derived a novel result (proposition 2.1) that allows us to obtain extended Laplace-Runge-Lenz vectors in a straightforward way for a quantum system of certain form. We have presented a new family of superintegrable systems which contain the one introduced previously in \cite{RW} as a member. We have obtained the exact solutions of the new systems by means of separation of variables in polar coordinates. We have obtained the integrals of motion of the systems and found the quadratic algebraic relations satisfied by these integrals. 

The quadratic algebra relations (\ref{qa1}), (\ref{qa2}) and (\ref{qa3}) found in section 5 for our model (\ref{sys}) is useful. They can be applied to algebraically obtain the energy spectrum of (\ref{sys}) by an approach similar to that used in \cite{LMZ} to find the spectrum for the special $N=D$ case (\ref{N=D}). Let us emphasis that the higher rank quadratic algebra and in particular the quadratic subalgebras generated by $\{Y_{p},Z_{p}\}$, $p \in \{2,...,N\}$, have a very interesting form: the structure constants involve polynomials of the Casimir operators of higher rank Lie algebras $so(d_p)$. Such quadratic algebra structures are non-trivial deformations of the $so(D+1)$ symmetry algebra of the Kepler-Coulomb system.

Our approach uses the Euler operator and block of variables which preserve the symmetry algebras $\oplus_{i} so(d_{i})$, and differs from other approaches such as co-algebras \cite{rod} and symplectic reduction scheme \cite{bal} that have been developed in context of classical superintegrable systems. This allows us not only to build new superintegrable systems, but also from a mathematical perspective to obtain new quadratic algebras and their explicit realizations. Classification, universal enveloping algebras and representation theory for such algebraic structures remain to be developed.

It is also worth pointing out that one can generalize our models by replacing the coupling constants $\alpha_i$ in \eqref{sys} by suitable functions of angles associated with the blocks $\mathcal{B}_i$. In such cases, the equations for the angular parts can be solved in terms of so-called $X_m$ Jacobi polynomials \cite{G1,G2,G3}.


\section*{Acknowledgements}

IM was supported by Australian Research Council Discovery Project DP 160101376. YZZ was  supported by
National Natural Science Foundation of China (Grant No. 11775177).




\begin{thebibliography}{}
\bibitem{1}
Fock, V. (1988). On the theory of the hydrogen atom. In: Dynamical Groups and Spectrum Generating Algebras. (In 2 Volumes) (pp. 400-410).
\bibitem{2}
Englefield, M. J., \& Biedenharn, L. C. (1974). Group theory and the Coulomb problem. Amer. J. Phys. 42, 263.
\bibitem{3}
Loebl, E. M. (Ed.). (2014). Group theory and its applications. Academic Press.
\bibitem{4}
Jauch, J. M., \& Hill, E. L. (1940). On the problem of degeneracy in quantum mechanics. Phys. Rev. 57, 641.
\bibitem{5}
Moshinsky, M. (1969). The harmonic oscillator in modern physics: from atoms to quarks. Gordon \& Breach.
\bibitem{6}
Miller Jr, W., Post, S., \& Winternitz, P. (2013). Classical and quantum superintegrability with applications. J. Phys. A: Math. Theor. 46, 423001.
\bibitem{7}
Kalnins, E. G., Kress, J. M., \& Miller Jr, W. (2013). Extended Kepler-Coulomb quantum superintegrable systems in three dimensions. J. Phys. A: Math. Theor. 46, 085206.
\bibitem{8}
Kalnins, E. G., Kress, J. M., \& Miller Jr, W. (2006). Second order superintegrable systems in conformally flat spaces. IV. The classical 3D St\"ackel transform and 3D classification theory. J. Math. Phys. 47, 043514.
\bibitem{9}
Kalnins, E. G., Miller, W., \& Post, S. (2010). Models for the 3D singular isotropic oscillator quadratic algebra. Phys. Atom. Nucl. 73, 359.
\bibitem{10}
Escobar-Ruiz, M. A., \& Miller Jr, W. (2017). Toward a classification of semidegenerate 3D superintegrable systems. J. Phys. A: Math. Theor. 50, 095203.
\bibitem{11}
Tanoudis, Y., \& Daskaloyannis, C. (2011). Algebraic calculation of the energy eigenvalues for the nondegenerate three-dimensional Kepler-Coulomb potential.  SIGMA 7, 054.
\bibitem{RW}
Rodrıguez, M. A., \& Winternitz, P. (2002). Quantum superintegrability and exact solvability in $n$ dimensions. J. Math. Phys. 43, 1309.
\bibitem{KWP}
Kalnins, E. G., Williams, G. C., Miller Jr, W., \& Pogosyan, G. S. (2002). On superintegrable symmetry-breaking potentials in $n$-dimensional Euclidean space. J. Phys. A: Math. Gen. 35, 4755.
\bibitem{Evan}
Evans, N. W. (1990). Superintegrability in classical mechanics. Phys. Rev. A, 41, 5666.
Evans, N. W. (1991). Group theory of the Smorodinsky-Winternitz system. J. Math. Phys. 32, 3369.
\bibitem{GPS}
Grosche, C., Pogosyan, G. S., \& Sissakian, A. N. (1995). Path integral discussion for Smorodinsky-Winternitz potentials: I. Two and three dimensional Euclidean space. Fortschr. Phys. 43, 453.
\bibitem{KWP1}
Kalnins, E. G., Williams, G. C., Miller Jr, W., \& Pogosyan, G. S. (1999). Superintegrability in three-dimensional Euclidean space. J. Math. Phys. 40, 708.
\bibitem{LMZ}
Liao, Y., Marquette, I., \& Zhang, Y.-Z. (2018). Quantum superintegrable system with a novel chain structure of quadratic algebras. J. Phys. A: Math. Theor. 51, 255201.
\bibitem{G1}
G\'omez-Ullate, D., Kamran, N., \& Milson, R. (2009). An extended class of orthogonal polynomials defined by a Sturm-Liouville problem. J. Math. Anal. Appl. 359, 352.
\bibitem{G2}
G\'omez-Ullate, D., Kamran, N., \& Milson, R. (2010). Exceptional orthogonal polynomials and the Darboux transformation. J. Phys. A: Math. Theor. 43, 434016.
\bibitem{G3}
G\'omez-Ullate, D., Kamran, N., \& Milson, R. (2012). On orthogonal polynomials spanning a non-standard flag. Contemp.  Math. 563, 51.
\bibitem{rod}
Rodriguez, M.A., Tempesta P., \& Winternitz P. (2008). Reduction of superintegrable systems the anisotropic harmonic oscillator, Phys. Rev. E78 046608
\bibitem{bal}
Ballesteros, A., Herraz F.F., \& Ragnisco O. (2008). Superintegrability on sl(2)-coalgebra spaces, Phys.of Atom. Nuclei 71 812 
\end{thebibliography}
\end{document}